\documentstyle[11pt,paspconf,epsf]{article}

\begin{document}

\title{The Role of Starburst in the Chemical Evolution of Galaxies}

\author{R. Coziol}
\affil{Divis\~ao de Astrofisica, Instituto Nacional de Pesquisas Espaciais, \\
CP 515, 12201-970, S.J.Campos - SP, BRAZIL}

\begin{abstract}
In a diagram of metallicity (\~z) vs. luminosity (M$_B$), the different types of
nearby (Z $< 0.05$) starburst galaxies seem to follow
the same linear relationship as the normal spiral and irregular galaxies. However,
for comparable luminosities the more massive starburst nucleus galaxies (SBNGs) show a
slight metallic defficiency as compared to the giant spiral galaxies. Furthermore, the  
SBNGs do not seem to follow the same relationship between \~z and Hubble type (T) than
the normal galaxies. The early-type SBNGs are metal poor as compared to normal galaxies.
It may suggests that the chemical evolution of a majority of the nearby starbursts galaxies
is not completely over and that the present burst represent an important phase of this process.
\end{abstract}

\keywords{starburst galaxies, chemical evolution, galaxy formation, star formation}

\section{Metallicity--luminosity relation for starbursts}

Garnett \& Shields (1987) demonstrated that spiral and irregular galaxies
display a metallicity (\~z) luminosity (M$_B$) relationship
over a wide range of magnitudes. 
In a recent paper Zaritsky, Kennicutt and Huchra (1994; hereafter ZKH) have shown that 
the individual \~z vs. M$_B$ relationships among spirals and irregulars merge to form a 
correlation that spans over 10 magnitudes in M$_B$ and over a factor of more than 100 in \~z. 
They have further argued that this correlation is the same as the one found 
for the elliptical and dwarf spheroidal galaxies (Brodie \& Huchra 1992).
Finally, ZHK have shown that there also
exists a strong correlation between \~z and the Hubble type (T).
They concluded that 
the abundance properties of the galaxies are imprimed early in their
evolution, and are related to the same initial conditions that
determine the Hubble types, the gas fractions and the
bulge-to-disk ratios. 

Following ZHK, the observed correlations of \~z with mass and Hubble types 
suggest that stochastic effects, such as starburst, do not affect the global abundance
of most galaxies. 
Contrary to this point of view, we show in Fig. 1 that the two types
of starburst galaxies (the HII galaxies and the starburst nucleus galaxies, SBNGs) 
seem to follow the same linear relationship between \~z and M$_B$ as 
the irregular and normal spiral galaxies (see Coziol 1996b, for a complete discussion). 

\begin{figure}[h]
\plotfiddle{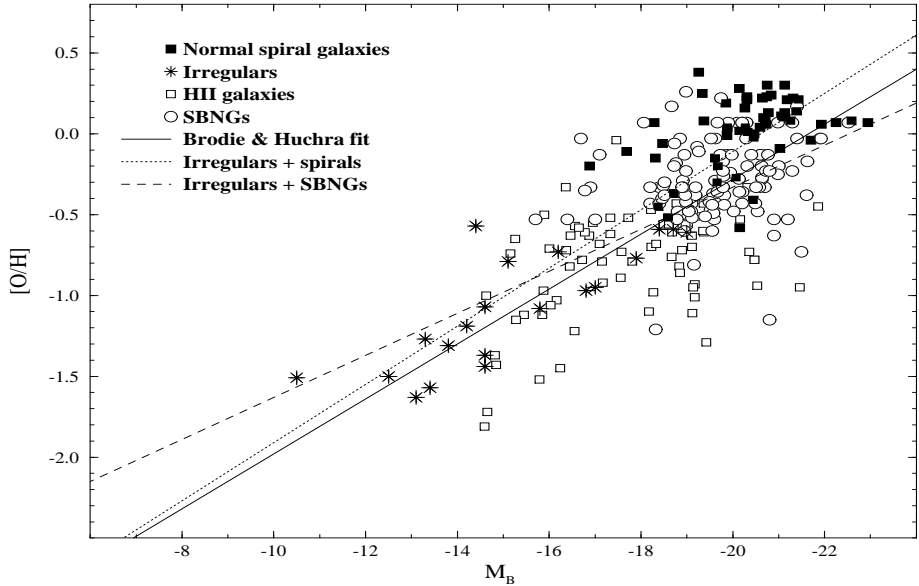}{2.8in}{270}{60}{50}{-220}{265}
\caption{Relationship between z and M$_B$ for the two types of starburst
galaxies as compared to the irregular and normal spiral galaxies.
The metallicity is given by [O/H] = log(O/H)$_{gal}$ - log(O/H)$_\odot$.
The giant spirals and irregular galaxies follow a slightly steeper relation 
than the elliptical and the dSph, while the relationship is flatter for 
the starburst galaxies. 
} 
\end{figure}

In Fig. 1, it is suggested that for comparable luminosities the SBNGs 
are slightly defficient in metals as compared to the giant spiral galaxies.
In Fig. 2, the few SBNGs galaxies in our sample
with a well defined Hubble type are plotted in a \~z vs. T diagram.  
Although the number of galaxies is small, it suggests that 
it is mostly the early-type SBNGs that are defficient in metals.

The lower metallicity of the SBNGs is incompatible with the usual interpretation
of these objects as a brief increase of star formation 
in a well evolved galaxy (Huchra 1977). On the contrary, the above result suggests 
that the SBNGs are less chemically evolved than normal galaxies. 
Consequently, the present burst of star formation could have 
a significant impact on the chemical evolution of these galaxies.
The fact that the starbursts follow a metallicity--luminosity relation 
suggests also some underlying regulating mechanism.
Starburst event is not such a chaotic phenomenon after all. 

\section{Discussion}

What is the cause of the metallicity--luminosity relation for the starbursts, and
why are the massive early-type SBNGs less metal rich than normal galaxies?
Four different hypotheses are considered. All suppose
a metallicity--mass relationship.

Outflows from SNe winds 
could produce a metallicity--mass relationship. 
However, such mechanism should be less efficient in massive galaxies,
unless we consider superwinds (Heckman, Armus \& Miley 1987). Following this model, 
a SNe rate of 3--30 yr$^{-1}$ could produce outflows of 
10--100 M$_{\odot}$  yr$^{-1}$.
If a typical starburst duration is 10$^8$ yr, a massive galaxy could
lose up to 10$^9$--10$^{10}$ M$_{\odot}$ of enriched matter. 
But, such a violent event seems to fit  
only one type of starbursts i.e. the ultra-luminous infrared galaxies.
It is also difficult to understand how such a catastrophic phenomenon could produce
a thight metallicity--mass relation without destroying the host galaxy. 

\begin{figure}[h]
\plotfiddle{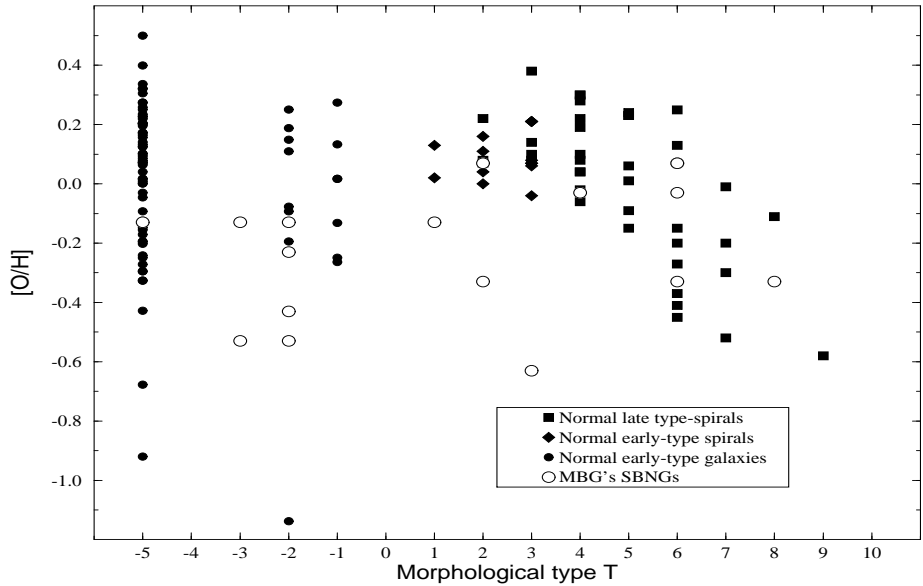}{2.8in}{270}{60}{50}{-220}{265}
\caption{Relationship between \~z and Hubble type (T) for the normal spiral galaxies
and the SBNGs. 
For the normal early-type galaxies the metallicity really corresponds to the metallicity
of the stars, which is given by [Fe/H] (see ZHK for a description of how the
metallicities were evaluated).} \label{fig-2}
\end{figure}

In principle, massive gas accretion of unprocessed material could diminished the
mean metallicity of normal evolved galaxies.  
Recent accretion is obvious in the SBNGs which posssess 
unusually high gas fraction for their morphological type
(Coziol {\em et al.} 1995).
The presence of some heavy elements also implies star formation, which
is expected as new accretion of gas will probably be followed by a new phase of star formation. 
However, the bursts of star formation have also to
last a sufficiently long time in order to modify significantly the chemical composition of the gas. 
We should imagine also an effective mechanism to bring all this gas in the nucleus.
It seems likely that we need a very fine tuning between the different parameters of this model 
to produce a tight metallicity--mass relation. Finally, it is difficult to imagine 
how huge reservoirs of gas could still exist in the neighbourhood of massive galaxies.

Tinsley \& Larson (1979) have shown that if elliptical 
galaxies and the bulges of spirals 
are formed by a gradual merging of small mass elements,
a metallicity--mass relationship is produced assuming that
the efficiency of the star formation increases with the mass of the merged systems. 
Following this scenario, a galaxy will also
acquire a disk if after the violent merger phase there are still residual 
outlying gas or gas--rich small subsystems that continue to be accreted. 
Following Struck--Marcell (1981), this phenomenon produce a higher yield, and therefore 
could explain the steeper linear relation observed for the giant spirals. 
Following this hypothesis, one would have to admit that
most of the nearby starbursts are still in the process of forming their bulges. 
Furthermore, if the SBNGs are results of mergers, we should observe the parent population of
these merging galaxies sometime in the recent past. 

Following the Stochastic Self--Propagating Star Formation (SSPSF)
theory (Gerola, Seiden \& Schulman 1980) the average rate of star formation increases 
with the size of the galaxy and that produces am etallicity--mass relationship. 
This is because in a large system, the probability 
that the star formation will percolate into areas which are fertiles
increases with the size of the galaxy. Massive galaxies will therefore experienced more bursts
of star formation and consequently looks more evolved than small mass galaxies.
Following this theory, the starburst are regulated by one internal
mechanism with a time duration longer 
than the usual dynamical time scale of interacting galaxies. 
Furthermore, all the starbursts are explained by the same mechanism. 
   
\section{Conclusion} 
  
Of the four scenarios considered above, only 
superwinds and massive accretion could save the ``old ''  or evolved nature of the 
host galaxies of the SBNGs. The chaotic nature of the first and the
fine tuning of the other seem difficult to reconciliate with a tight metallicity--mass
relation. It remains either the multiple mergers scenario
or the SSPSF model. These two models suggest that the chemical evolution 
of galaxies proceeds through a sequence of bursts of star formation. 
Traces of such sequences may still be detectable in the nearby SBNGs (Coziol 1996a).
Both models also suggests that the starburst phenomenon 
is really a fundamental process of the chemical evolution of galaxies.

\acknowledgments

I would like to thank J. E. Steiner for discussing part of this article 
and Hugo V. Capelato for its critical reading.
The financial
support of the brazilian FAPESP ({\em Funda\c{c}\~ao de Amparo \`a
Pesquisa do Estado de S\~ao Paulo}), under contract 94/3005--0 is
gratefully acknowledged.

\end{document}